# 基於安全偽隨機數產生器的安全密碼產生器

Abel C. H. Chen
*Information & Communications Security Laboratory,*
*Chunghwa Telecom Laboratories*
Taoyuan, Taiwan
ORCID: 0000-0003-3628-3033

*摘要*—近幾年持續有報導指出有網站的帳號和文字型密碼(以下簡稱「文字型密碼」為「密碼」)外洩，導致有可能影響個人資料洩露的風險，反應出資訊安全和密碼安全的重要性。然而，造成帳號和密碼外洩的原因，除了網站安全之外，密碼本身的安全性也是重要的。因此，如何制定一個足夠安全的密碼，將也是建構整體安全性和保護個人資料的重要環節之一。有鑑於此，本研究提出基於安全偽隨機數產生器(Pseudo Random Number Generator, PRNG)的安全密碼產生器，結合雜湊訊息鑑別碼(Keyed-Hash Message Authentication Code, HMAC)、密文訊息鑑別碼(Cipher-based Message Authentication Code, CMAC)、或 KECCAK 訊息鑑別碼(KECCAK Message Authentication Code, KMAC)建構偽隨機數產生器產製安全的隨機數，再運用安全的偽隨機數產製密碼。為驗證本研究提出的方法產製的隨機數，在實驗中參考國家標準暨技術研究院(National Institute of Standards and Technology, NIST)特別出版(Special Publication, SP) 800-90B 分別從熵(Entropy)驗證和獨立且同分布(Independent and Identically Distributed, IID)驗證兩個面向進行實證。並且，由實驗結果顯示，本研究提出的方法通過熵驗證和獨立且同分布驗證，產製的密碼具備足夠的隨機性和安全性。

*關鍵字*—*偽隨機數產生器、KMAC、密碼產生器、熵驗證、獨立且同分布驗證*

## I. 前言

近年來持續有駭客通過多種不同的攻擊手法竊取大量網路平台的帳號和文字型密碼(以下簡稱「文字型密碼」為「密碼」)，將可能導致嚴重的個人資料外洩的風險[1]。然而，造成帳號和密碼外洩的原因，除了網站本身的安全漏洞、釣魚攻擊[2]等之外，還有密碼本身的安全性。因此，有許多研究陸續整理不安全密碼的特性，包含有基於關鍵字的密碼[3]、基於熱門字詞的密碼[4]、基於使用者資訊的密碼[5]、密碼重覆使用[6]-[8]等。這些特性都將導致密碼的不安全，讓攻擊者容易猜測出密碼的文字組合。甚至近幾年在人工智慧和大型語言模型的快速發展下，攻擊者有可能結合人工智慧生成技術來猜測密碼，並且可以提升攻擊的成功率[9]-[10]。

有鑑於此，許多網路平台也陸續結合多種技巧來防範駭客攻擊，包含有結合一次性密碼[11]或結合認證機制[12]避免重試攻擊、結合雙因子認證[13]來提升安全性、以及密碼特殊加密後儲存在資料庫[14]等策略。然而，為建構足夠安全的帳號和密碼，除了用上述策略來防禦外，密碼本身的安全性也是需要提升的。近幾年，開始有研究在討論如何設計一個夠安全的密碼[15]。

因此，本研究提出基於安全偽隨機數產生器(Pseudo Random Number Generator, PRNG)的安全密碼產生器，結合雜湊訊息鑑別碼(Keyed-Hash Message Authentication Code, HMAC)[16]-[17]、密文訊息鑑別碼(Cipher-based Message Authentication Code, CMAC)[18]-[19]、或 KECCAK 訊息鑑別碼(KECCAK Message Authentication Code, KMAC)[20]-[21]建構偽隨機數產生器產製安全的隨機數，再運用安全的偽隨機數產製密碼。可以讓使用者可選的(optional)設定其偏好的文字作為待雜湊的訊息(To-Be-Hashed Message, TBHM)(即安全偽隨機數產生器的輸入)，由安全偽隨機數產生器產製足夠隨機的偽隨機數，再根據密碼可用字元集合和長度，由偽隨機數來產製文字型密碼。本研究的主要貢獻條列如下：

- 本研究建構基於雜湊訊息鑑別碼、基於密文訊息鑑別碼、基於 KECCAK 訊息鑑別碼的安全偽隨機數產生器，提出基於安全偽隨機數產生器的安全密碼產生器，通過安全偽隨機數產生器產製安全的偽隨機數，再運用安全的偽隨機數產製密碼。

- 本研究分析在不同密碼可用字元集合和長度下的安全性，並且對比該密碼與進階加密標準-128 (Advanced Encryption Standard-128, AES-128) 和 AES-256 [19]的安全性。

- 參考美國國家標準暨技術研究院(National Institute of Standards and Technology, NIST) 出版的 SP (Special Publication) 800-90B [22]分別從熵(Entropy)驗證和獨立且同分布(Independent and Identically Distributed, IID)驗證兩個面向進行實證。

本論文總共包含五個章節。第 II 節主要介紹偽隨機數產生器和驗證隨機性的方法。第 III 節詳細說明本研究提出的基於安全偽隨機數產生器的安全密碼產生器，並且討論在不同密碼可用字元集合和長度限制下的安全性。第 IV 節描述實驗環境，並且分別驗證隨機性和計算效率。最後，第 V 節總結本研究發現和討論未來發展方向。

## II. 文獻探討

第 II.A 節和第 II.B 節將先分別介紹基於線性同餘產生器的偽隨機數產生器和安全偽隨機數產生器，說明偽隨機數產生器的原理和作法。第 II.C 節將說明驗證隨機數的方法，並且說明熵驗證和獨立且同分布驗證。

### A. 基於線性同餘產生器的偽隨機數產生器

由於 C 程式語言和 Java 程式語言等知名的程式語言和套件內建的 Random 類別都是建構在線性同餘產生器的基礎上[23]，所以本節將先說明基於線性同餘產生器的偽隨機數產生器的原理及其運作方式。

為提供偽隨機數，基於線性同餘產生器的偽隨機數產生器可以指定隨機數種子(seed) $k$，當隨機數種子一樣時，則產製出來的隨機數也將一樣。如果沒有指定隨機數種子，則預設的作法則是先讀系統資訊和系統時間來作為隨機數種子，再執行同樣的演算法來產製隨機數。公式(1)表示當代入隨機數種子 $k$ 後，與乘數 $a$ 進行邏輯互斥或(Exclusive-OR) $\oplus$ 計算，再模整數 $m$ 得到的初始值 $f_0(k)$。

$$f_0(k) \equiv a \oplus k \pmod{m}. \tag{1}$$

當欲產製生第 $i$ 個隨機數 $f_i(k)$ 時，假設隨機數種子為 $k$，並且第 $i-1$ 個隨機數是 $f_{i-1}(k)$，運用公式(2)把 $f_{i-1}(k)$ 乘上乘數 $a$ 再加上一增量整數 $c$，再模整數 $m$ 限制隨機數長度得到 $f_i(k)$。

$$f_i(k) \equiv a \times f_{i-1}(k) + c \pmod{m}. \tag{2}$$

然而，需要注意的是基於線性同餘產生器的偽隨機數產生器並不安全，從公式(2)可以觀察到可以運用第 $i$ 個隨機數 $f_i(k)$ 推導出第 $i-1$ 個隨機數是 $f_{i-1}(k)$，如公式(3)所示；依此類推，可以運用第 $i$ 個隨機數 $f_i(k)$ 一直回推出隨機數種子。因此，在線上運作的系統，不建議採用基於線性同餘產生器的偽隨機數產生器，避免存在偽隨機數被攻擊的風險。

$$f_{i-1}(k) \equiv (f_i(k) - c) \times a^{-1} \pmod{m}. \tag{3}$$

### B. 安全偽隨機數產生器

有鑑於基於線性同餘產生器的偽隨機數產生器不安全，陸續有安全偽隨機數產生器被提出和被開發出來，美國國家標準暨技術研究院出版 SP 800-108 Rev. 1，內容整理了基於雜湊訊息鑑別碼、基於密文訊息鑑別碼、基於KECCAK 訊息鑑別碼等方法可以產製安全的偽隨機數[21]。本節將分別說明基於雜湊訊息鑑別碼、基於密文訊息鑑別碼、基於 KECCAK 訊息鑑別碼的設計及其原理。

#### 1) 基於雜湊訊息鑑別碼

基於雜湊訊息鑑別碼的安全性主要建構在雜湊函數不可逆的特性，可以結合安全雜湊演算法 2 (Secure Hash Algorithm-2, SHA2)[17]和安全雜湊演算法 3 (Secure Hash Algorithm-3, SHA3)[20]來提供訊息鑑別碼。其中，輸入變數有已經過前處理後的金鑰值 $k$ (金鑰位元長度為雜湊函數區塊(block)的位元長度 $l_h$)和訊息值 $M$，通過與填充值 ipad (重覆'00110110'位元直到雜湊函數區塊的位元長度 $l_h$)和 opad (重覆'01011100'位元直到雜湊函數區塊的位元長度 $l_h$)進行邏輯互斥或(Exclusive-OR) $\oplus$ 計算，並且選擇合適的雜湊函數 $h$，運用公式(4)計算得到訊息鑑別碼 $r_{hmac}(k, M)$[16]。

$$r_{hmac}(k, M) = h\big((k \oplus opad) || h((k \oplus ipad) || M)\big). \tag{4}$$

由於基於雜湊訊息鑑別碼每次產製的訊息鑑別碼長度為 $l_h$ bits，所以欲產製長度 $L$ bits 的隨機數時，則需要產製 $\left\lceil \frac{L}{l_h} \right\rceil$ 個訊息鑑別碼。為產製 $\left\lceil \frac{L}{l_h} \right\rceil$ 個不同的訊息鑑別碼，本研究參考 NIST SP 800-108 Rev. 1 設計的計數器模式(Counter Mode)[21]運用公式(5)修改訊息 $M$ 為 $M_i$，再代入 $r_{hmac}(k, M_i)$ 以產製第 $i$ 個訊息鑑別碼(如公式(6)所示)。

$$M_i = i || KDF || 0x00 || M || L. \tag{5}$$

$$r_{hmac}(k, M_i) = h\big((k \oplus opad) || h((k \oplus ipad) || M_i)\big) = r_{hmac,i}. \tag{6}$$

#### 2) 基於密文訊息鑑別碼

基於密文訊息鑑別碼的安全性主要建構在進階加密標準的安全性，可以結合 AES-128 和 AES-256 [19]並且採用密文區塊鏈模式(Chiper Block Chaining Mode, CBC Mode)來提供訊息鑑別碼。其中，輸入變數有已經過前處理後的金鑰值 $k$ (金鑰位元長度為進階加密標準區塊的位元長度 $l_a$)和訊息值 $M$，並且根據在進階加密標準區塊的位元長度 $l_a$ 對切割成 $\left\lceil \frac{Length(M)}{l_a} \right\rceil$ 個區塊(如公式(7)所示)。之後根據密文區塊鏈模式的計算方法，把第 $i$ 個區塊密文和第 $i+1$ 區塊明文進行邏輯互斥或(Exclusive-OR) $\oplus$ 計算，然後再用金鑰值 $k$ 執行 AES 函數得到第 $i+1$ 區塊密文(如公式(8)所示)。直到第 $\left\lceil \frac{Length(M)}{l_a} \right\rceil$ 個區塊運用公式(9)計算得到訊息鑑別碼 $r_{cmac}(k, Split(M))$[18]。其中，Pad0 函數表示對訊息往右邊填充 0 直到達到進階加密標準區塊的位元長度 $l_a$，Pad1 函數表示對訊息往右邊填充 1 直到達到進階加密標準區塊的位元長度 $l_a$。

$$Split(M) = M'_1 || M'_2 || \ldots || M'_{\left\lceil \frac{Length(M)}{l_a} \right\rceil}. \tag{7}$$

$$c_{i+1} = AES(k, c_i \oplus M'_{i+1}),$$
$$\text{where } c_0 = Pad0(0) \text{ and } 0 \leq i < \left\lceil \frac{Length(M)}{l_a} \right\rceil. \tag{8}$$

$$r_{cmac}(k, Split(M))$$
$$= AES\left(k, c_{\left\lceil \frac{Length(M)}{l_a} \right\rceil - 1} \oplus M^*_{\left\lceil \frac{Length(M)}{l_a} \right\rceil}\right),$$
$$\text{where } K_2(k) = AES(k, Pad(0)) \ll 2 \text{ and}$$
$$M^*_{\left\lceil \frac{Length(M)}{l_a} \right\rceil} = Pad1\left(M'_{\left\lceil \frac{Length(M)}{l_a} \right\rceil}\right) \oplus K_2(k). \tag{9}$$

由於基於密文訊息鑑別碼每次產製的訊息鑑別碼長度為 $l_a$ bits，所以欲產製長度 $L$ bits 的隨機數時，則需要產製 $\left\lceil \frac{L}{l_a} \right\rceil$ 個訊息鑑別碼。為產製 $\left\lceil \frac{L}{l_a} \right\rceil$ 個不同的訊息鑑別碼，本研究參考 NIST SP 800-108 Rev. 1 設計的計數器模式運用公式 (5) 修改訊息 $M$ 為 $M_i$，再代入 $r_{cmac}(k, Split(M_i))$ 以產製第 $i$ 個訊息鑑別碼(如公式(10)所示)。

$$r_{cmac}(k, Split(M_i))$$
$$= AES\left(k, c_{i, \left\lceil \frac{Length(M)}{l_a} \right\rceil - 1} \oplus M^*_{i, \left\lceil \frac{Length(M)}{l_a} \right\rceil}\right). \tag{10}$$

#### 3) 基於 KECCAK 訊息鑑別碼

基於 KECCAK 訊息鑑別碼的安全性主要建構在安全雜湊演算法 KECCAK (Secure Hash Algorithm KECCAK, SHAKE)雜湊函數不可逆的特性[20]來提供訊息鑑別碼。其中，美國國家標準暨技術研究院在安全雜湊演算法 KECCAK 基礎上設計可客製化安全雜湊演算法 KECCAK (customizable Secure Hash Algorithm Keccak, cSHAKE)方法。輸入變數有已經過前處理後的金鑰值 $k$ (金鑰位元長度為雜湊函數區塊的位元長度 $l_k$)、訊息值 $M$、輸出長度

$L$，通過公式(11)把訊息值 $M$ 修改為 $M^*$，然後再搭配字串 $S$ 代入可客製化安全雜湊演算法 KECCAK (如公式(12)所示)計算得到訊息鑑別碼 $r_{kmac}(k,M,L,S)$[20]。除此之外，可客製化安全雜湊演算法 KECCAK 有另一個特色是可以設定輸出長度 $L$，所以欲產製長度 $L$ bits 的隨機數時，可以在呼叫可客製化安全雜湊演算法 KECCAK 時直接設定即可產製符合需求長度的隨機數。

$$M^* = k||M||L. \quad (11)$$

$$\begin{aligned} r_{kmac}(k,M,L,S) \\ = CSHAKE(M^*, L, \text{"KMAC"}, S). \end{aligned} \quad (12)$$

### C. 隨機數的隨機性驗證

為證明隨機數的隨機性，美國國家標準暨技術研究院出版 SP 800-90B，內容指出熵驗證和獨立且同分布驗證的相關規範。本節將分別說明驗證方法和相關門檻值。

#### 1) 熵驗證

為驗證隨機數的熵，本研究參考 NIST SP 800-90B 所述驗證方法[22]建立 1000×1000 大小的 restart 矩陣，運用隨機數產生器每次產製 1000 個隨機位元值，並且重新啟動 1000 次。其中，第 $i$ 次第 $j$ 個隨機位元值是 restart 矩陣第 $i$ 列第 $j$ 欄元素值。再計算 restart 矩陣的每一行和每一列的 0 和 1 發生的次數，以及取得最高次數作為 Most Common Value (MCV)，再根據 MCV 換算為最小熵 (mini_entropy)和運用二項式分布檢定得到 $p$-value。並且根據 NIST SP 800-90B 定義的 $p$-value 門檻值為 0.000005 [22]，進行驗證是否高於門檻值。

#### 2) 獨立且同分布驗證

為驗證隨機數的獨立且同分布，本研究參考 NIST SP 800-90B 所述驗證方法，分別從獨立性檢定(Independence Test, Ind. Test)、適應度檢定(Goodness-of-fit Test, GF Test)、以及最長重覆子字串長度檢定(Length of the Longest Repeated Substring Test, Length of the LRS Test)[22]三個面向進行驗證，分述如下。

(1). 獨立性檢定(Ind. Test)：把 restart 矩陣每一列資料(即 1000 個隨機位元值)各別切割為 10 等份，每一等份裡各有 100 個隨機位元值，所以如果 0 和 1 在均勻分布下期望值為各 50 次。再依據真值和期望值計算出卡方值和 $p$-value，並且根據 NIST SP 800-90B 定義的 $p$-value 門檻值為 0.001 [22]，進行驗證是否高於門檻值。

(2). 適應度檢定(GF Test)：把 restart 矩陣每一列資料(即 1000 個隨機位元值)各別切割為 500 等份，每一等份裡各有 2 個隨機位元值(即 00~11 的 4 種組合之一)，所以如果 00~11 在均勻分布下每一種組合的期望值為各 125 次。再依據真值和期望值計算出卡方值和 $p$-value，並且根據 NIST SP 800-90B 定義的 $p$-value 門檻值為 0.001 [22]，進行驗證是否高於門檻值。

(3). 最長重覆子字串長度檢定(Length of the LRS Test)：觀察 restart 矩陣每一列資料(即 1000 個隨機位元值)中最長重覆子字串的長度，再運用二項式分布檢定計算 $p$-value。其中，$p$-value 門檻值在 NIST SP 800-90B [22]中定義為 0.001，所以在長度為 1000 時對應的最長重覆子字串的長度門檻值大約為 28。

## III. 本研究提出的基於 KMAC 偽隨機數產生器的安全密碼產生器

第 III.A 節先介紹本研究提出的的基於安全偽隨機數產生器的安全密碼產生器的設計。第 III.B 節討論密碼可用字元集合與密碼長度及其安全性。

### A. 設計理念

本研究設計的基於安全偽隨機數產生器的安全密碼產生器整體流程如圖 1 所示。其中，根據密碼規則，可以決定可用字元集合 $Q$ 大小為 $q$，並且決定密碼長度為 $n$。為提升安全性，本研究為隨機產製的密碼中的每個字元是從 $N$-bit 長度的隨機數映射產製，也就是從值域$[0, 2^N – 1]$映射到值域$[0, q – 1]$。因此，本研究設計的基於安全偽隨機數產生器的安全密碼產生器需先由偽隨機數產生器產製 $nN$ 個隨機位元值(即 $L = nN$) $T$，再依序切割成 $n$ 個 $N$-bit 長度的隨機數(即第 $i$ 個隨機數 $t_i$ 是由第$(iN + 1)$個位元值到第$((i + 1)N)$個隨機位元值組成)(如公式(13)所示)。之後再運用公式(14)把第 $i$ 個隨機數映射為第 $i$ 個密碼字元 $p_i$，再把每個密碼字元連接(concatenate)在一起即可產製密碼 $P$(如公式(15)所示)。當 $N$ 值越大，則越安全，但相對計算時間會更長。在本研究將採用第 II 節介紹的偽隨機數產生器作為核心的偽隨機數產生器，用以產製 $nN$ 個隨機位元值。

$$Split(T) = t_1||t_2||...||t_n. \quad (13)$$

$$p_i = t_i \pmod{q}. \quad (14)$$

$$P = p_1||p_2||...||p_n. \quad (15)$$

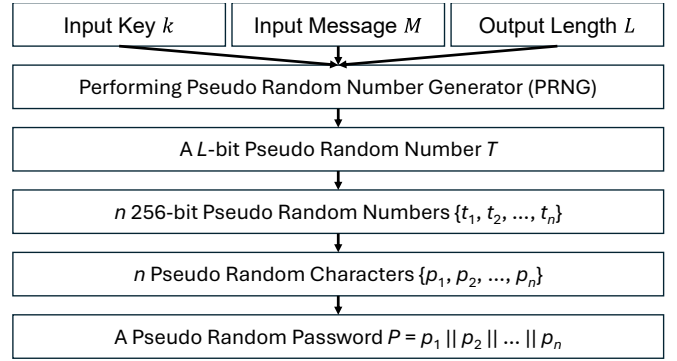

Fig. 1. 本研究提出的基於 KMAC 偽隨機數產生器的安全密碼產生器

其中，在呼叫安全偽隨機數產生器時，需代入 3 個參數：金鑰值 $k$、訊息值 $M$、輸出長度 $L$。在金鑰值 $k$ 的部分，可以採用系統當下時間或是使用者指定時間，並且由於時間在電腦裡可能是存成 long 資料型態(只有 64 bits)，可以採用多次雜湊函數計算來產製金鑰值 $k$ (符合區塊長度)。訊息值 $M$ 為使用者自行指定的字串，可以是自己原本所熟悉的密碼字串。輸出長度 $L$ 為 $nN$。通過前述設定後即可產製符合使用者指定內容產製的安全偽隨機數，並且在該基礎上產製安全的密碼。

### B. 密碼可用字元集合與密碼長度

本節首先將先定義密碼可用字元集合，再討論密碼長度和安全性分析。

## 1) 密碼可用字元集合

密碼可用字元集合主要取決於網路平台的設定，不同的網路平台可能採用不同的密碼可用字元集合，所以本研究採用下面種密碼可用字元集合進行討論。後續將分別從不同密碼可用字元集合深入討論和比較。

- 密碼可用字元集合 1：僅採用英文字母小寫和英文字母大寫，即 $Q_1$ = {'a', 'b', …, 'z', 'A','B', …, 'Z'}。因此，可用字元集合 $Q_1$ 大小為 $q_1 = 52$。
- 密碼可用字元集合 2：$Q_1$ 和數字，即 $Q_2$ = {$Q_1$, '0','1', …, '9'}。因此，可用字元集合 $Q_2$ 大小為 $q_2 = 62$。
- 密碼可用字元集合 3：$Q_2$ 和特殊字元，即 $Q_3$ = {$Q_2$, '~','!', '@', '#', '$', '%', '^', '+', '-', '='}。因此，可用字元集合 $Q_3$ 大小為 $q_3 = 72$。

## 2) 密碼長度及其安全性分析

本節以 AES-128 安全強度的情況下，著重討論在不同密碼可用字元集合的情況下適合的密碼長度 $n$。其中，由於 AES-128 主要建構在 128 bits 的 0 或 1 的均勻分佈下，所以可用字元集合 $Q$ 大小為 $q$ 時每個字元約為 $\log_2 q$ bits，所以 128 bits 需要 $\frac{128}{\log_2 q}$ 個字元以上。因此，對比 AES-128 安全強度，每種密碼可用字元集合所對應的密碼長度如表 I 所示。例如，當採用英文字母小寫、英文字母大寫、數字、特殊字元的集合(即密碼可用字元集合3)時，仍需要密碼長度仍應達 21 個字元以上才等價於 AES-128 安全強度，並且是建構在均勻分佈下。如果密碼字元組成不是均勻分佈，則可能不夠安全。

TABLE I. 各種密碼可用字元集合所需的密碼長度

| 密碼可用字元集合 | 集合大小 | 密碼長度 $n$ |
|---|---|---|
| $Q_1$ | $q_1 = 52$ | $\lceil \frac{128}{\log_2 52} \rceil = \lceil 22.45 \rceil = 23$ |
| $Q_2$ | $q_2 = 62$ | $\lceil \frac{128}{\log_2 62} \rceil = \lceil 21.50 \rceil = 22$ |
| $Q_3$ | $q_3 = 72$ | $\lceil \frac{128}{\log_2 72} \rceil = \lceil 20.75 \rceil = 21$ |

## IV. 實驗結果與討論

本節將先介紹實驗環境，然後分別討論隨機性驗證結果、密碼字元分布比較結果、計算時間比較結果。

### A. 實驗環境

本研究比較 4 種偽隨機數產生器，包含基於線性同餘產生器的偽隨機數產生器、基於雜湊訊息鑑別碼的偽隨機數產生器、基於密文訊息鑑別碼的偽隨機數產生器、基於 KECCAK 訊息鑑別碼的偽隨機數產生器，分別從分析其安全性和計算效率。

由於本研究主要採用 Java 進行實作，所以基於線性同餘產生器是建構在 Java 內建的 Random 類別。另外，本研究採用 BouncyCastle 1.80 實作雜湊訊息鑑別碼、密文訊息鑑別碼、KECCAK 訊息鑑別碼。硬體規格為 Intel(R) Core(TM) i7-10510U CPU、16 GB RAM。

### B. 隨機性驗證結果

在驗證偽隨機數的隨機性，本研究參考 NIST SP 800-90B 所述驗證方法[22]分別從熵驗證和獨立且同分布驗證進行實證，詳細設定可參考本文的第 II.C 節。

圖 2 為各種偽隨機數產生器的熵驗證的 $p$-value。由實驗結果可以觀察到每一種偽隨機數產生器都能產製足夠亂度(熵值足夠大)的隨機數，並且皆高於 NIST SP 800-90B 規範的 $p$-value 門檻值為 0.000005。因此，每種偽隨機數產生器皆通過熵驗證，並且基於雜湊訊息鑑別碼的偽隨機數產生器、基於密文訊息鑑別碼的偽隨機數產生器、基於 KECCAK 訊息鑑別碼的偽隨機數產生器的 $p$-value 更高，表示這些偽隨機數產生器產製的隨機數更亂、更安全。

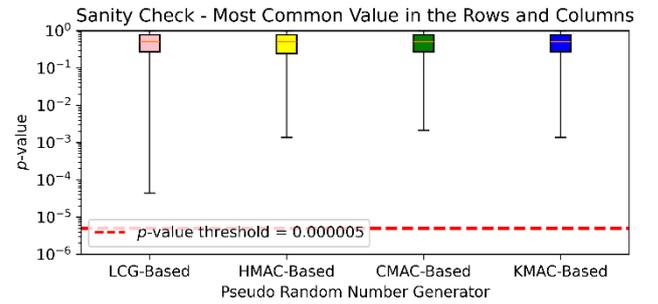

Fig. 2. 熵驗證

表 II 為各種偽隨機數產生器的獨立且同分布驗證的結果，分別呈現獨立性檢定、適應度檢定、最長重覆子字串長度檢定的 $p$-value 中位數。由實驗結果顯示，每個偽隨機數產生器的 $p$-value 皆高於 NIST SP 800-90B 規範的 $p$-value 門檻值為 0.001。因此，絕大部分下都符合獨立且同分布。

TABLE II. 獨立且同分布驗證 $P$-VALUE 中位數

| PRNG | Ind. Test | GF Test | Length of the LRS Test |
|---|---|---|---|
| LCG-Based | 0.401 | 0.285 | 1.000 |
| HMAC-Based | 0.398 | 0.288 | 1.000 |
| CMAC-Based | 0.391 | 0.310 | 1.000 |
| KMAC-Based | 0.398 | 0.294 | 1.000 |

### C. 密碼字元分布比較結果

本節主要比較運用各種偽隨機數產生器搭配本研究第 III 節所提出的方法產製的密碼字元是否符合均勻分布，並且運用卡方檢定來驗證；如果 $p$-value 高於 0.01 則表示與均勻分布無顯著差異，密碼亂度夠高。本研究在各種密碼可用字元集合限制下，基於各種偽隨機數產生器各別產製 10000 次密碼，再進行統計檢定。表 III 為各種偽隨機數產生器在各種密碼可用字元集合限制下的結果，由實驗結果可以觀察到皆能提供足夠亂度的密碼。

TABLE III. 密碼字元分布卡方檢定結果 $P$-VALUE

| PRNG | $Q_1$ | $Q_2$ | $Q_3$ |
|---|---|---|---|
| LCG-Based | 0.444 | 0.390 | 0.982 |
| HMAC-Based | 0.968 | 0.600 | 0.909 |
| CMAC-Based | 0.593 | 0.912 | 0.197 |
| KMAC-Based | 0.682 | 0.933 | 0.437 |

## D. 計算時間比較結果

本節主要比較運用各種偽隨機數產生器搭配本研究第 III 節所提出的方法產製密碼的計算時間。本研究在各種密碼可用字元集合限制下，基於各種偽隨機數產生器各別產製 10000 次密碼，運用盒鬚圖的方式呈現實驗結果，分別如圖 3、圖 4、以及圖 5 所示。由實驗結果可以觀察到在不同的密碼可用字元集合限制下計算時間上並無顯著差異，並且都是呈現基於線性同餘產生器的偽隨機數產生器有最高的效率，而基於 KECCAK 訊息鑑別碼的偽隨機數產生器的效率次之。

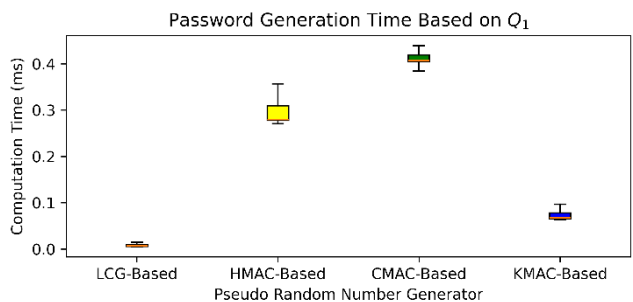

Fig. 3. 密碼可用字元集合 1 的各種密碼產生器計算時間比較結果

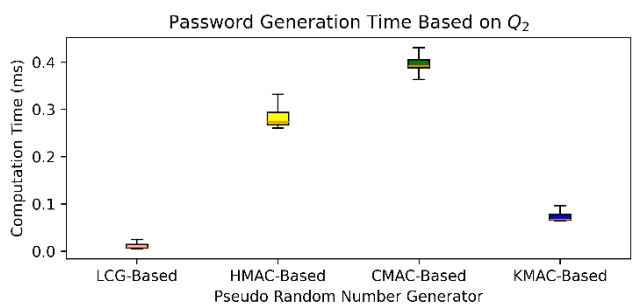

Fig. 4. 密碼可用字元集合 2 的各種密碼產生器計算時間比較結果

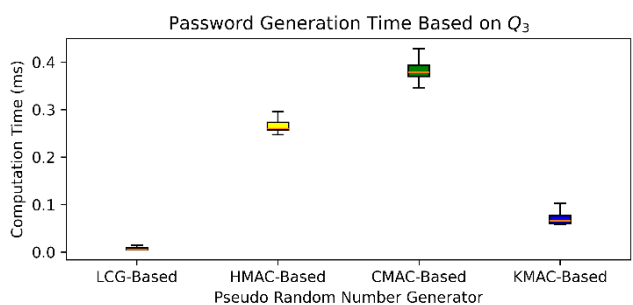

Fig. 5. 密碼可用字元集合 3 的各種密碼產生器計算時間比較結果

## E. 小結與討論

由第 IV.B 節和第 IV.C 節的實驗結果表示各種偽隨機產生器都能產製足夠隨機性的隨機數，並且密碼字元也能服從均勻分布。在第 IV.D 節的實驗結果表示基於線性同餘產生器的偽隨機數產生器的計算時間最短，而基於 KECCAK 訊息鑑別碼的偽隨機數產生器的計算時間次之。

值得注意的是，雖然實驗結果表示基於線性同餘產生器的偽隨機數產生器可以產製足夠隨機性的隨機數，並且有較高的效率。然而，線性同餘產生器具有可逆的特性，所以將會被破解。另外，由於 KECCAK 訊息鑑別碼是建構在雜湊函數的基礎上，具有不可逆的特性，所以可以提供更安的偽隨機數。因此，建議未來主要可以採用基於 KECCAK 訊息鑑別碼的偽隨機數產生器的安全密碼產生器。

## V. 結論與未來研究

本研究提出基於安全偽隨機數產生器的安全密碼產生器，並且分別探索 4 種偽隨機數產生器，包含基於線性同餘產生器的偽隨機數產生器、基於雜湊訊息鑑別碼的偽隨機數產生器、基於密文訊息鑑別碼的偽隨機數產生器、基於 KECCAK 訊息鑑別碼的偽隨機數產生器。在第 III.B 節用數學模型和理論證明在各種密碼可用字元集合限制下適合的密碼長度。在第 IV 節分別運用實作證明 4 種偽隨機數產生器產製的偽隨機數的隨機性，並且證明在這些偽隨機數產生器產製的密碼的隨機數。以及比較這些偽隨機數產生器產製密碼的計算時間。最後，根據實驗結果綜合評選出合適的安全密碼產生器，建議採用基於 KECCAK 訊息鑑別碼的偽隨機數產生器的安全密碼產生器。

在未來研究中，可以考慮結合量子隨機數產生器，產製真隨機數來提供更安全的安全密碼產生器。因此，在未來量子隨機數產生器晶片可以部署到個人電腦、智慧型手機等，再運用本研究提出的基於安全偽隨機數產生器的安全密碼產生器進行結合，即可提供更安全的密碼產生器。